\documentclass [preprint,prb,amsmath,amssymb,a4paper,showpacs] {revtex4}

\usepackage{graphicx}% Include figure files
\usepackage{dcolumn}% Align table columns on decimal point
\usepackage{bm}% bold math

\begin{document}

%\preprint{APS/}

\title{Dielectric Properties of Noncrystalline HfSiON}%

\author{Masahiro Koike}
% \altaffiliation[Also at ]{Physics Department, XYZ University.}%Lines break automatically or can be forced with \\
 \email{m-koike@amc.toshiba.co.jp}
\author{Tsunehiro Ino}%
\author{Yuuichi Kamimuta}%
\author{Masato Koyama}%
\author{Yoshiki Kamata}%
\author{Masamichi Suzuki}%
\author{Yuichiro Mitani}%
\author{Akira Nishiyama}%

\affiliation{%
Corporate Research and Development Center, Toshiba Corporation,\\ 
5 Shinsugita-cho, Isogo-ku, Yokohama 235-8522, Japan
}%

\date{\today}% It is always \today, today,
             %  but any date may be explicitly specified

\begin{abstract}
The dielectric properties of noncrystalline hafnium silicon oxynitride (HfSiON) films with a variety of atomic compositions were investigated. The films were deposited by reactive sputtering of Hf and Si in an O, N, and Ar mixture ambient. The bonding states, band-gap energies, atomic compositions, and crystallinities were confirmed by X-ray photoelectron spectroscopy (XPS), reflection electron energy loss spectroscopy (REELS), Rutherford backscattering spectrometry (RBS), and X-ray diffractometry (XRD), respectively. The optical (high-frequency) dielectric constants were optically determined by the square of the reflective indexes measured by ellipsometry. The static dielectric constants were electrically estimated by the capacitance of Au/HfSiON/Si(100) structures. It was observed that low N incorporation in the films led to the formation of only Si-N bonds without Hf-N bonds. An abrupt decrease in band-gap energies was observed at atomic compositions corresponding to the boundary where Hf-N bonds start to form. By combining the data for the atomic concentrations and bonding states, we found that HfSiON can be regarded as a pseudo-quaternary alloy consisting of four insulating components: SiO$_2$, HfO$_2$, Si$_3$N$_4$, and Hf$_3$N$_4$. The optical and static dielectric constants for the films showed a nonlinear dependence on the N concentration, whose behavior can be understood in terms of abrupt Hf-N bond formation. 

\end{abstract}

\pacs{77.22.-d, 77.55.+f, 61.43.-j}% PACS, the Physics and Astronomy
                             % Classification Scheme.
% 77.22.-d Dielectric properties of solids and liquids 
% 77.55.+f Dielectric thin films
% 61.43.-j Disordered solids (see also 81.05.Gc, 81.05.Kf, and 81.05.Rm in materials science; for photoluminescence of disordered solids, see 78.55.Mb and 78.55.Qr) 
% 

%\keywords{Suggested keywords}%Use showkeys class option if keyword
                              %display desired
\maketitle

\section{\label{sec:level1}INTRODUCTION}

Downsizing of electric devices has been aggressively pursued to realize large-scale integrated circuits (LSIs) with sub-50-nm-technology nodes in the near future. During this process, the gate insulator film, which is one of the key components in LSIs, will become less than 2 nm thick, \cite{ITRS04} and SiO$_2$, which has been used for the gate insulator, will reach a fundamental physical limit. \cite{schulz99} Such ultra-thin films required in the future are unable to sufficiently suppress the leakage current, resulting in high electric power consumption. This is due mainly to the quantum tunneling effect of electrons through the film. Since this effect is inevitable in principle, new insulators to replace SiO$_2$ will be needed to achieve further reductions in thickness. 

High-dielectric constant (high-k) materials have recently attracted a great deal of attention as an alternative gate insulator because the use of gate insulators with a higher dielectric constant can reduce the electrical thickness without a reduction in the physical thickness. \cite{kingon00} On the other hand, the insulator must have a wide band gap energy that corresponds to the barrier height, the band off-set between the insulator and the electrode, to reduce the tunneling currents from the electrodes. Since materials with higher dielectric constants tend to have a narrower band-gap energy (e.g., Ref. \onlinecite{ziman72}), and the tunneling current is strongly dependent on both the traveling distance and the barrier height, materials with an excessively high dielectric constants such as TiO$_{2}$ and Ta$_{2}$O$_{5}$ are not suitable for use as the gate insulator. Thus, an ideal insulator should have a sufficiently high band-gap energy and a moderately high dielectric constant. A large number of high-k materials such as metal oxides tend to react with silicon substrates to generate an interfacial layer with a low dielectric constant between the material and the substrate. \cite{hubbard96} Since the formation of such a layer with a low dielectric constant leads to an increase in the effective thickness, a film that is stable on silicon is required. In addition, it is desirable that gate insulators manufactured by conventional processes for LSIs remain in the noncrystalline phase even after annealing at high temperature, typically 1,000$\char'27\kern-.3em\hbox{C}$, to electrically activate impurities such as B and As in polycrystalline silicon gate electrodes and source/drain regions in transistors. 

Among many high-permittivity materials, HfSiON is a possible alternative to SiO$_{2}$ for the future generation of LSIs. This material satisfies all of the above conditions, e.g., the films are able to maintain the noncrystalline phase even at 1,000$\char'27\kern-.3em\hbox{C}$. \cite{visokay03} In addition, the dielectric properties of Hf- and Zr- based high-k materials are a topic of growing interest in the field of material science as well as semiconductor technology. The dielectric constants of Hf and Zr silicates and oxides have been studied theoretically \cite{rignanese02,fiorentini02,zhao02} and experimentally.\cite{lucovsky00,wilk001,wilk002} It has been found that the static dielectric constant of Zr silicate (ZrSiO) shows a supralinear dependence on the Zr concentration. \cite{lucovsky00,wilk001,wilk002} On the other hand, a nearly linear dependence on the concentration has also been shown by calculation. \cite{rignanese02} Although there have been several reports concerning the dielectric properties of Zr-based high-permittivity materials, few studies, particularly experimental studies, have investigated the dielectric characteristics of Hf-based high-k materials such as HfSiON. In addition, there is interest regarding how Hf-N bonds affect the properties of HfSiON. However, this has yet to be fully clarified.

In the present study, we fabricated HfSiON films with various atomic compositions and systematically investigated their properties such as bonding states, band-gap energies, atomic compositions, and basic dielectric constants. The results showed that HfSiON films behave as an insulator, even though the films contain a large number of Hf-N bonds, that the dielectric constants have a nonlinear dependence on the N concentration, and that these properties can be explained by the bonding states and atomic compositions.

\section{EXPERIMENTAL}

We fabricated HfSiON with various atomic compositions and relatively thick films ($\sim$100 nm). These films were deposited on $p$-type Si(100) wafers with co-sputtering of Hf and Si targets in an O$_{2}$, N$_{2}$, and Ar mixture ambient at room temperature. The wafers were treated with HF to remove native silicon oxide on the surface before deposition. During deposition, the atomic compositions were precisely controlled: the nitrogen atomic concentration $C_{\rm N}$ was regulated by adjusting the flow rates of Ar, O$_{2}$, and N$_{2}$. The ratio of the hafnium atomic concentration to the total metal concentration, the $C_{\rm Hf}/(C_{\rm Hf}+C_{\rm Si})$ ratio, was varied by changing the power ratio imposed on Hf to Si targets. The atomic compositions were estimated by Rutherford backscattering spectrometry (RBS). The film thicknesses were confirmed by both ellipsometry and scanning electron microscopy (SEM). Ellipsometry analysis was also used to examine the reflective index, n, of the films. It was found by X-ray diffractometry (XRD) measurement using Cu K$\rm \alpha$ radiation that all of the films remained in the noncrystalline phase. The static dielectric constants were measured using capacitors with Au electrode/HfSiON($\sim$100 nm)/$p$-Si(100) substrate/Al back contact structures. Au and Al were deposited by thermal vaporization. The Au gate electrode was formed on the films through a metal shadow mask with round holes that controlled the size and shape of the area. The static dielectric constants of the films were measured by the capacitance-voltage ($C$-$V$) method \cite{nicollianl82} at 10 kHz, a frequency that is sufficiently low to estimate the static dielectric constant. The static dielectric constant, $\varepsilon_{S}$, was extracted from the capacitance using the relation: $C = \varepsilon_{S} \varepsilon_{0} A / T_{\rm phys}$, where $A$ is the area of the capacitor and $T_{\rm phys}$ is the physical thickness. The optical dielectric constants, $\varepsilon_{\infty}$, were extracted from the square of the reflective index, $n$. Since the frequency of a laser in ellipsometry is on the order of 100 THz, only electrons can follow the changing electric field at such high frequencies. Reflection electron energy loss spectroscopy (REELS) was used to determine the band gap energies for the HfSiON films, although X-ray photoelectron spectroscopy (XPS) is widely employed to estimate band-gap energies. This is because the fact that Hf $4s$ peaks exist very close to the energy loss peak of O $1s$ poses difficulties in dealing with these films using XPS. The bonding states in these films, especially Si $2p$ and Hf $4f$ spectra, were analyzed by XPS. Monochromatic Al K$\rm \alpha$ was used as the X-ray source. The photoelectrons were detected from the sample normal. The spectra were corrected by the C $1s$ peak from C absorption on the surface to eliminate the shift due to the charging up of the films. XPS and REELS measurements were carried out using a modified version of a commercially available instrument (AXIS-ULTRA, Kratos Analytical Ltd.).

\section{RESULTS and DISCUSSION}

\subsection{Bonding states and band-gap energies}\label{sec3a}

\begin{figure}
\includegraphics[width=16cm]{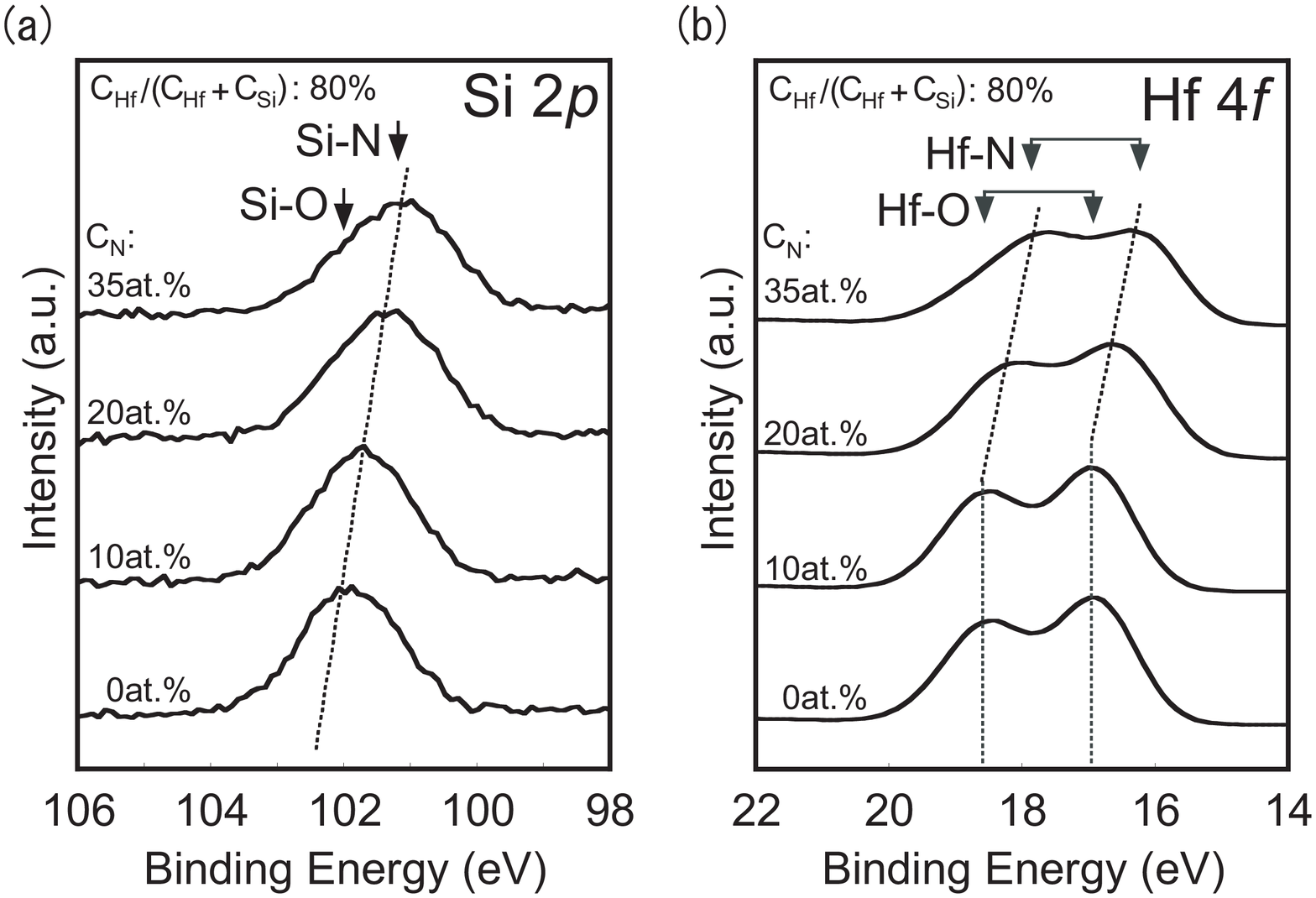}% Here is how to import EPS art
\caption{\label{fig1} XPS spectra for (a) Si $2p$ and (b) Hf $4f$ of HfSiON films with a $C_{\rm Hf}/(C_{\rm Hf}+C_{\rm Si})$ ratio of 80 \% and $C_{\rm N}$ values of 0, 10, 20, and 35 at.\%. The Si $2p$ and Hf $4f$ peaks shift to lower binding energies as $C_{\rm N}$ increases, meaning that Si-O and Hf-O bonds decrease whereas Si-N and Hf-N bonds increase. The Hf $4f$ peaks remain at the same energy up to a $C_{\rm N}$ of 10 at.\% and start to move toward a lower binding energy above 20 at.\%. This suggests the preferential formation of Si-N and Hf-O bonds rather than Hf-N and Si-O bonds.}
\end{figure}

Figures \ref{fig1} (a) and \ref{fig1} (b) show typical examples of Si $2p$ and Hf $4f$ peaks for the films, respectively. The $C_{\rm Hf}/(C_{\rm Hf}+C_{\rm Si})$ ratio is $\sim$80 \% , and the $C_{\rm N}$ values are 0, $\sim$10, $\sim$20, and $\sim$35 atomic \% (at.\% ). 

In the Si\ $2p$ spectra (Fig. \ref{fig1} (a)), we see a single peak around 101-102 eV. The peak in this range of binding energies originates from Si-O and/or Si-N bonds, and the ratio of the mixture of the two determines the position of the peak. The peak of Si $2p$ for HfSiO appears at 102 eV, which is lower than that for SiO$_{2}$ ($\sim$104 eV). This can be explained by the difference in the next nearest neighbor (NNN) atom between them; \cite{opilal02,guittet01} the nearest atoms of Si for both SiO$_{2}$ and HfSiO (i.e., C$_{\rm N}$=0 at.\%) are O. On the other hand, the NNN of Si in SiO$_{2}$ is always Si whereas that in HfSiO is either Si or Hf, depending on the ratio of Si to Hf. Since Hf has electron-donating characteristics, the charge transferred to O from NNN Hf is more than that from NNN Si. The increase in charges transferred from NNN to O causes an increase in the electron density of Si, thereby decreasing the binding energy of Si $2p$. 

\begin{figure}
\includegraphics[width=10cm]{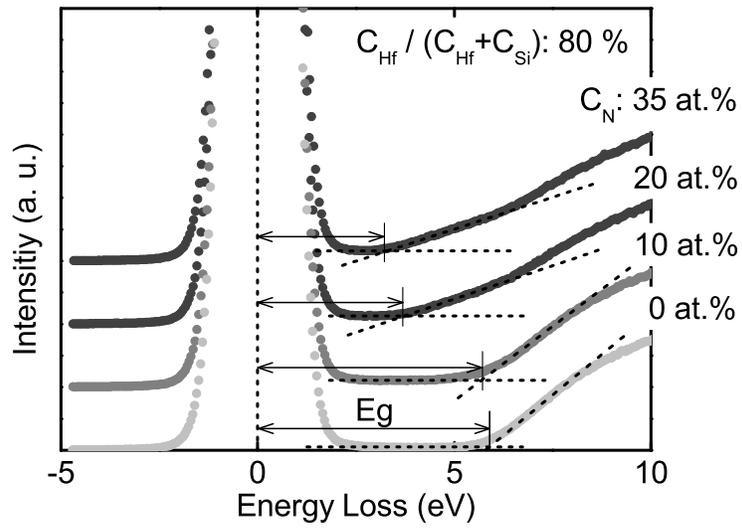}% Here is how to import EPS art
\caption{\label{fig2} REELS spectra for HfSiON films with a $C_{\rm Hf}/(C_{\rm Hf}+C_{\rm Si})$ ratio of 80 \% and $C_{\rm N}$ values of 0, 10, 20, and 35 at.\%. The distance between the center of a peak and the position of the rise in the spectrum corresponds to the band-gap energy. The band-energy tends to become narrower as $C_{\rm N}$ in the film increases.}
\end{figure}

The peak moves to a lower binding energy as $C_{\rm N}$ increases from 0 to 35 at.\%. This behavior indicates that Si-O bonds decrease and Si-N bonds increase as $C_{\rm N}$ increases. The Hf $4f$ peak shows a doublet corresponding to the spin-orbit coupling for Hf $4f_{5/2}$ and Hf $4f_{7/2}$ at lower and higher binding energies, respectively. Like Si $2p$ spectra, Hf $4f$ peaks also shift to a lower binding energy as $C_{\rm N}$ increases. This general behavior in our HfSiON films is the same as that in HfON films that were previously reported; \cite{kang02} however, it should be noted that there is a specific difference in behaviors between Si $2p$ and Hf $4f$ peaks: in the case of Hf $4f$ spectra, peaks stay at the same energy in the range of small amounts of $C_{\rm N}$ ($<$20 at.\%), and start to shift from 20 at.\% of $C_{\rm N}$. This suggests that N tends to bond to Si as compared to Hf in HfSiON. From another point of view, we could say that O tends to bond to Hf as compared to Si. It should be noted that no peaks are visible around 99 eV in Si $2p$ or around 14 eV in Hf $4f$, meaning that there are no Si-Si, Hf-Hf, or Hf-Si bonds in the films. In addition, we confirmed that almost no peaks corresponding to N-O bonds appeared in N $1s$ or O $1s$ spectra. These results indicate that the films are composed almost entirely of bonds of pairs of cations and anions, i.e., Si-O, Si-N, Hf-O, and Hf-N. Since SiO$_{2}$, Si$_{3}$N$_{4}$, and HfO$_{2}$ are well-known as insulators, HfSiON films composed of Si-O, Si-N, and Hf-O bonds should operate as insulators, but films with Hf-N bonds as well should be conductive because HfN is well-known as a metal with a rock salt structure like NaCl. \cite{stampfl01} Therefore, we measured the band gap energies using REELS to confirm the effects of the bonding states, especially those of Hf-N bonds, on the electronic structures. Figure \ref{fig2} shows the REELS for the films, which were the same films as used for the XPS measurements. These intensities are normalized at the largest peaks around 0 eV, i.e., the elastic scattering peak. REELS is an analytic technique for detecting the energy of reflected electrons with a fixed primary energy. Electrons reflected from the sample lose their energies by plasmon oscillation or transition from the valance band to the conduction band (electron-hole interband excitation) or transition from the core level to the conduction band at the sample surface. Among these possible energy loss processes, we conclude that these onsets are attributable to the electron-hole excitation because the values for the samples without nitrogen agreed with the optical band gap energies estimated previously.\cite{kato02} As illustrated in Fig. \ref{fig2}, the band gap energies tend to decrease as $C_{\rm N}$ increases and show abrupt changes between 10 and 20 at.\%. This $C_{\rm N}$ level corresponds to the composition at which Hf-N bond formation starts (Fig. \ref{fig1}). It should be noted that HfSiON with a $C_{\rm N}$ of more than 20 at.\% shows $\sim$3 eV band gap energies even though a large number of Hf-N bonds exist. Thus, HfSiON films containing Hf-N bonds appear to exhibit insulating behavior.

\subsection{Alloy compositions}

\begin{figure}
\includegraphics[width=10cm]{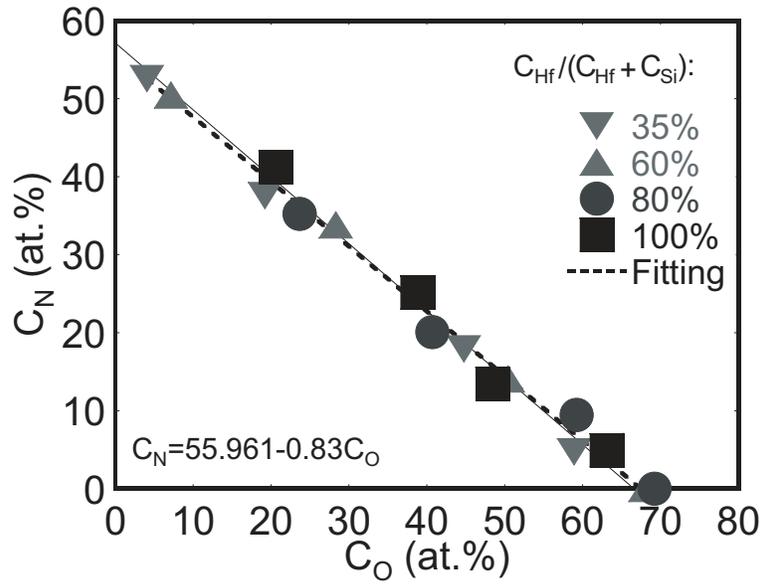}% Here is how to import EPS art
\caption{\label{fig3} Relationship between $C_{\rm N}$ and $C_{\rm O}$ in HfSiON films with various atomic compositions. The atomic compositions were obtained by RBS. All data fall close to a single line that satisfies the relation: $2 C_{\rm O} + 3 C_{\rm N} = 4 (C_{\rm Si} + C_{\rm Hf})$. From this relation and the bonding state obtained from XPS data, HfSiON can be regarded as a pseudo-quaternary alloy: [(SiO$_{2}$)$_{x}$(HfO$_{2}$)$_{1-x}$]$_{z}$ [(Si$_{3}$N$_{4}$)$_{y}$ (Hf$_{3}$N$_{4}$)$_{1-y}$]$_{1-z}$, where "pseudo" means that the films are noncrystalline and that four atoms randomly connect with each other in this case. Since Hf$_{3}$N$_{4}$ is an insulator, it is probable that a pseudo-alloy including a Hf$_{3}$N$_{4}$ component shows insulating characteristics. This is consistent with the results for the band-gap energies.}
\end{figure}

Although the alloy structure of a crystalline substance is usually determined by X-ray diffraction (XRD) or similar techniques, it is impossible to directly determine the alloy composition of materials such as HfSiON films by such studies because they are noncrystalline. We therefore determined the alloy composition indirectly by the use of a combination of data for the bonding states obtained by XPS and the atomic compositions obtained by RBS. Figure \ref{fig3} shows the relationship between $C_{\rm O}$ and $C_{\rm N}$ of HfSiON with various atomic compositions. We found that all data points fell close to a single straight line that satisfies the following relation:
\begin{equation}
2 C_{\rm O} + 3 C_{\rm N} = 4 (C_{\rm Si} + C_{\rm Hf}),\label{eq1}
\end{equation}
where the total of the atomic compositions is normalized. Relation (\ref{eq1}) indicates the balance of the atomic concentrations of anions (left-hand side) and cations (right-hand side) and shows that the films satisfy charge neutrality. Before moving on, it is useful to consider the relationship in the special case of $C_{\rm Hf}=0$, i.e.,
\begin{equation}
2 C_{\rm O} + 3 C_{\rm N} = 4 C_{\rm Si}.\label{eq2}
\end{equation}
The atomic compositions of stoichiometric SiON satisfy relation (\ref{eq2}). This reflects the fact that SiON forms a pseudo-binary alloy: \cite{hattangady96}
\begin{equation}
\rm (SiO_2)_{x} (Si_3N_4)_{1-x}.\label{eq3}
\end{equation}
Here, ``pseudo'' means that SiON consists of Si-O and Si-N bonds that are randomly connected. O is twofold coordinated and bonds to two Si, N is threefold coordinated and bonds to three Si, and Si is fourfold coordinated and bonds to four atoms of O and/or N. By analogy to the case of SiON, the pseudo-alloy of HfSiON can be determined similarly. From the combination of the bonding states obtained by XPS, which demonstrate the existence of only anion-cation bonds, and the atomic compositions obtained by RBS, which satisfy relation (\ref{eq1}), HfSiON can be regarded as a pseudo-quaternary alloy:
\begin{equation}
\rm [(SiO_{2})_{x} (HfO_{2})_{1-x} ]_{z} [(Si_{3}N_{4})_{y} (Hf_{3}N_{4})_{1-y}]_{1-z}.\label{eq4}
\end{equation}
It should be noted that Hf and N form Hf$_{3}$N$_{4}$ in this pseudo-alloy. Hf$_{3}$N$_{4}$ has a closed shell structure and is an insulating material according to the calculation by Kroll. \cite{kroll03,kroll04} Therefore, HfSiON alloy composed of four insulating units should be an insulator. We can figure that $C_{\rm O}$ and $C_{\rm N}$ determine z in (\ref{eq4}), while x and y in (\ref{eq4}) can vary if they satisfy equation (\ref{eq1}). In other words, HfSiON with an identical atomic composition could assume different bonding states that should lead to different band-gaps; however, XPS and REELS results always show the same bonding states and band-gap energies, respectively, for HfSiON with an atomic composition. We consider that this is due to the tendency for Si-N bonds to form more easily than Hf-N bonds, as discussed in section \ref{sec3a}. Thus, x and y in (\ref{eq4}) are uniquely determined by the atomic composition. As mentioned above, no Hf-N bonds in HfSiON films with a $C_{\rm Hf}/(C_{\rm Hf}+C_{\rm Si})$ ratio of 80 \% are observed at small $C_{\rm N}$ ($\leq$10 at.\%) and start to form when $C_{\rm N}$ exceeds 20 at.\%. Let us assume that N always bonds to Si first even if Hf exists, and bonds to Hf second after all Si has bonded to N. In other words, we can say that O always bonds to Hf first and to Si second. In this case, the concentration at which Hf-N bonds start to form corresponds to x=0, y=1, and z=12/13 in (\ref{eq4}), i.e., Hf$_{12}$O$_{24}$Si$_{3}$N$_{4}$. This simple calculation shows that 4/43, i.e., about 9.3 at.\% is the $C_{\rm N}$ level of the starting point. This is consistent with the changes in the bonding states (Fig. \ref{fig1}) and the band-gap energies (Fig. \ref{fig2}) in films with a $C_{\rm N}$ of around 10 at.\%. 

The electric structures of HfON and HfSiON have recently been calculated, \cite{shang04} although only crystals were treated. They assumed the same HfSiON alloy as (\ref{eq4}) and HfON alloys of x=0 and y=0 in (\ref{eq4}), i.e., (HfO$_{2}$)$_{z}$(Hf$_{3}$N$_{4}$)$_{1-z}$. Their results showed that the conduction band minima of both HfO$_{2}$ and HfON originate in Hf5d, and that the valence band maximum in HfO$_{2}$ is O $2p$ while that in HfON is N $2p$, which is nearer the vacuum energy level. Although increasing N reduced the band-gap energy due to Hf-N bond formation, there was still a small band-gap in their calculation. This supports our result that HfSiON films with a large number of Hf-N bonds continue to exhibit insulating behavior. 

\subsection{Optical and static dielectric constants}

\begin{figure}
\includegraphics[width=16cm]{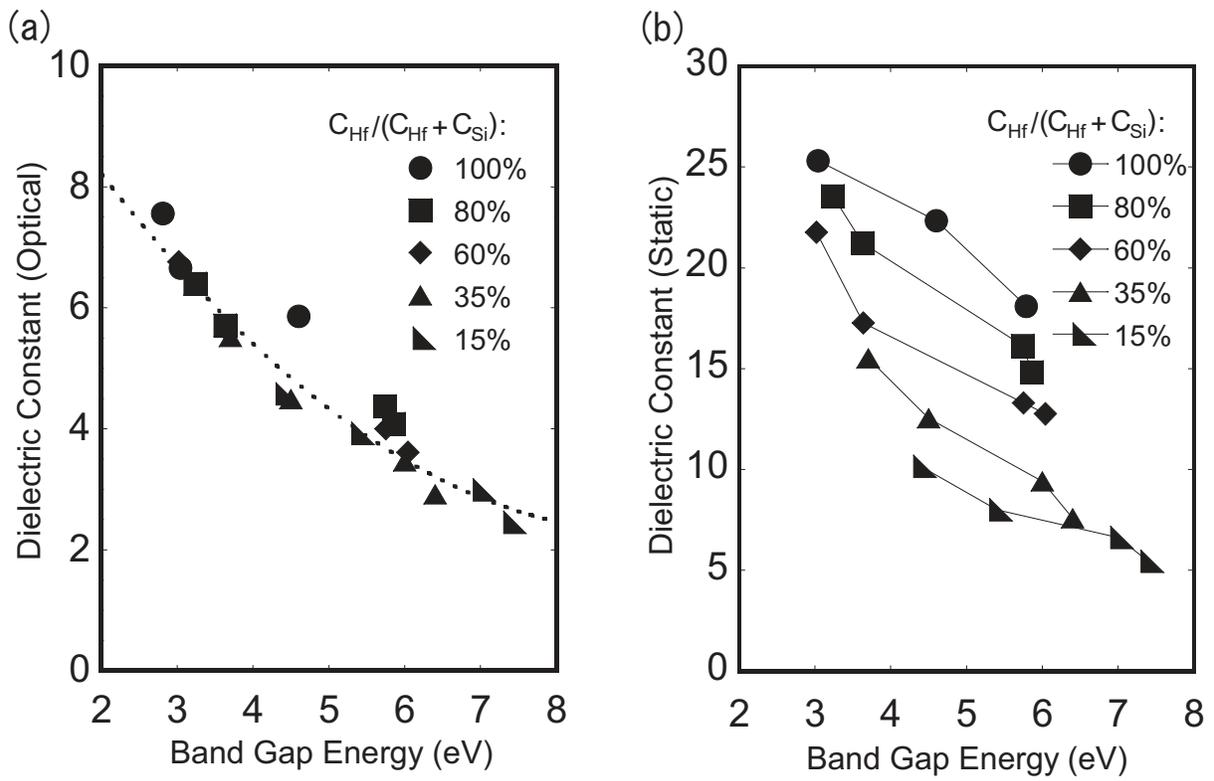}% Here is how to import EPS art
\caption{\label{fig4} Relationship between band-gap energy and dielectric constant. (a) Optical and (b) static dielectric constants. The optical dielectric constants were derived from the square of the refractive indexes measured by ellipsometry. The static dielectric constants were estimated by Capacitance-Voltage measurement at 10 kHz using capacitors with Au/HfSiON/$p$-Si(100) structures.}
\end{figure}

Figure \ref{fig4} (a) shows the relationship between the optical dielectric constant and the band-gap energy. Almost all of the data points fall very near the universal curve despite the variations in atomic composition, and the optical dielectric constants decrease with the band gap energy (Fig. \ref{fig4} (a)). This is due to the fact that the optical dielectric constant originating from electronic polarizability is strongly dependent on the width of the band gap energy, \cite{ziman72} and is nearly independent of the origin of the band gap, i.e., the atomic composition. On the other hand, the data for the static dielectric constant is different from that for the optical dielectric constant. The dielectric constant at the same band-gap energy tends to increase with the $C_{\rm Hf}/(C_{\rm Hf}+C_{\rm Si})$ ratio. This is because while the optical dielectric constant originates from only electronic polarization, \cite{ziman72} the static dielectric constant arises from both ionic and electronic polarization. A frequency of 10 kHz in $C$-$V$ measurements is so slow that ionic atoms as well as electrons can follow the change in the electric field. The increase in the $C_{\rm Hf}/(C_{\rm Hf}+C_{\rm Si})$ ratio makes the bonds in the films less covalent and more ionic, thereby causing an increase in the static dielectric constant even at the same band gap energy.

\begin{figure}
\includegraphics[width=16cm]{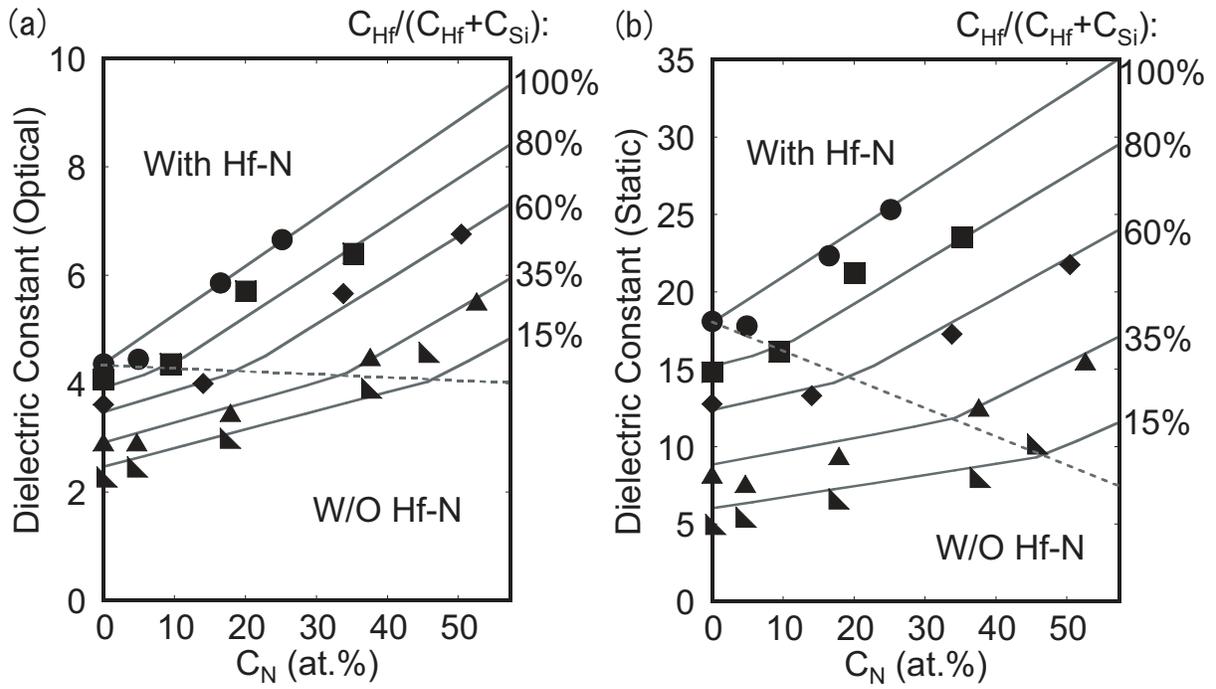}% Here is how to import EPS art
\caption{\label{fig5} Relationship between atomic composition and dielectric constant. (a) Optical and (b) static dielectric constants. The same symbols (experimental data) as in Fig. \ref{fig4} are used. The data were fitted using the relation: $\varepsilon = z [x \varepsilon_{\rm SiO_{2}} + (1-x) \varepsilon_{\rm HfO_{2}}] + (1-z) [y \varepsilon_{\rm Si_{3}N_{4}} + (1-y) \varepsilon_{\rm Hf_{3}N_{4}}]$, where $\varepsilon=\varepsilon_\infty (\varepsilon=\varepsilon_{S})$ for the optical (static) dielectric constant. The nonlinearity of the dielectric constants for each $C_{\rm Hf}/(C_{\rm Hf}+C_{\rm Si})$ ratio can be explained by the formation of Hf-N bonds.}
\end{figure}

Figures \ref{fig5} (a) and \ref{fig5} (b) show the optical and static dielectric constants, respectively, for HfSiON with various atomic compositions. Both of them have the same tendency to increase with $C_{\rm N}$. Interestingly, both of the dielectric constants fixed for the $C_{\rm Hf}/(C_{\rm Hf}+C_{\rm Si})$ ratio increase nonlinearly with $C_{\rm N}$. The shapes of the curves are concave.

We can understand this tendency as follows: assuming the HfSiON pseudo-alloy, the optical (static) dielectric constant of films with various compositions, $\varepsilon_\infty$ ($\varepsilon_{S}$), are modeled as a linear combination with the four optical (static) dielectric constants of the components in the alloy:
\begin{equation}
\varepsilon = z [x \varepsilon_{\rm SiO_{2}} + (1-x) \varepsilon_{\rm HfO_{2}}] + (1-z) [y \varepsilon_{\rm Si_{3}N_{4}} + (1-y) \varepsilon_{\rm Hf_{3}N_{4}}].\label{eq5}
\end{equation}
where $\varepsilon=\varepsilon_\infty\ (\varepsilon=\varepsilon_{S})$ for the optical (static) dielectric constant. This enables us to examine the behavior of the optical and static dielectric constants of HfSiON with various atomic compositions and bonding states if we know the dielectric constants of SiO$_{2}$, HfO$_{2}$, Si$_{3}$N$_{4}$, and Hf$_{3}$N$_{4}$. As is well known, the optical dielectric constants $\varepsilon_\infty \ (=n^{2})$ values for SiO$_{2}$ and Si$_{3}$N$_{4}$ are $\sim 2.13\ (=1.46^{2})$ and $\sim 4.0\ (=2.0^{2})$ while the static dielectric constants $\varepsilon_{S}$ are 3.9 and 7.4, respectively. The optical and static dielectric constants of HfO$_{2}$, which we can extract from our data (Fig. \ref{fig5} (a)), are $4.37\ (=2.09^{2})$ and 18, respectively. On the other hand, those of Hf$_{3}$N$_{4}$ could not be estimated directly. This is due to the difficulty in producing pure Hf$_{3}$N$_{4}$ in our experimental system because a tiny amount of residual oxygen leads to the oxidation of Hf atoms. We therefore estimated the dielectric constant by extrapolation of the line on which the data for HfON with $C_{\rm N}$ values in the range 0-35 at.\% are fitted by the least squares method. This resulted in optical and static dielectric constants of 9.5 and 35, respectively. There have been no reports concerning the optical dielectric constant for comparison with our data. On the other hand, a static dielectric constant of 30, which is close to our result, has been reported, with this value obtained for atomic layer deposited Hf$_{3}$N$_{4}$.\cite{becker04} In calculating the dielectric constant using relation (\ref{eq5}), we took into account the fact that N tends to bond to Si as compared to Hf. The results are shown in Fig. \ref{fig5} as solid lines that take the $C_{\rm Hf}/(C_{\rm Hf}+C_{\rm Si})$ ratios as the parameter. Also shown as a dotted line is the borderline that was extracted from our XPS data and that separates the regions with and without Hf-N bonds. The figure shows good agreement between the experimental data and the estimated lines. In this way, the nonlinearity of the dielectric constant as a function of $C_{\rm N}$ for each $C_{\rm Hf}/(C_{\rm Hf}+C_{\rm Si})$ ratio can be explained by the formation of Hf-N bonds. 

The driving force for the preferential formation of Si-N bonds compared to Hf-N bonds in HfSiON films has not yet been clarified. The difference in cohesive energy among SiO$_{2}$, HfO$_{2}$, Si$_{3}$N$_{4}$, and Hf$_{3}$N$_{4}$ is a possible explanation for this phenomenon; however, data for Hf$_{3}$N$_{4}$ has, to our knowledge, not yet been obtained. Further theoretical work as well as experiment work is needed to resolve this issue.

\section{CONCLUSION}

We studied noncrystalline HfSiON films with a variety of atomic concentrations, and obtained the following results. The bonding states from the XPS spectra indicated that Si-N bonds form preferentially compared to Hf-N bonds and that neither cation-cation nor anion-anion bonds are observed in the films. It was found by REELS measurements that the band gaps decrease as N and Hf concentrations increase and that an abrupt decrease is observed in accordance with Hf-N bond formation. Both optical and static dielectric constants are enhanced as the N and Hf concentrations increase. Whereas the optical dielectric constants relative to the band-gap energies fall close to a single curve at any $C_{\rm Hf}/(C_{\rm Hf}+C_{\rm Si})$ ratio and $C_{\rm N}$, the static dielectric constants at the band-gap energy increase with the $C_{\rm Hf}/(C_{\rm Hf}+C_{\rm Si})$ ratio. The effect of the preferential formation of Si-N bonds compared to Hf-N bonds also appears in both dielectric properties; the dielectric constant abruptly increases at the concentration where Hf-N bond formation starts. 

We found that the atomic compositions satisfy the relation, $2 C_{\rm O} + 3 C_{\rm N} = 4 (C_{\rm Si} + C_{\rm Hf})$ in units of atomic percent, indicating the charge neutrality of the films consisting of two cations (Si and Hf) and two anions (O and N). This relation and the bonding states revealed that this material is composed of a pseudo-quaternary alloy: $\rm [(SiO_{2})_{x}(HfO_{2})_{1-x}]_{z} [(Si_{3}N_{4})_{y}(Hf_{3}N_{4})_{1-y}]_{1-z}$. Since the pseudo-alloy includes not metallic Hf$_{1}$N$_{1}$ but insulating Hf$_{3}$N$_{4}$, HfSiON films that satisfy the alloy compositions consist of four insulating components and therefore have insulating properties. It was also found that the behavior of the dielectric constant could be explained by a linear combination of those of each element taking into account the tendency for bond formation.

\begin{acknowledgments}
The authors would like to thank M. Ono and A. Nara of Toshiba Corporation for their helpful comments and discussions.
\end{acknowledgments}

\newpage %Just because of unusual number of tables stacked at end
%\bibliography{PRB_HfSiON}% Produces the bibliography via BibTeX.

\newpage

\newpage

\newpage

\newpage

\newpage

\end{document}